\documentclass{sig-alternate-2013}

\usepackage{graphicx}

\usepackage{amsmath,latexsym}
\usepackage{wasysym}
\usepackage{multirow}
\usepackage{colortbl}
\usepackage{upgreek}
\usepackage{listings}
\usepackage{color}
\usepackage{url}

\newfont{\mycrnotice}{ptmr8t at 7pt}
\newfont{\myconfname}{ptmri8t at 7pt}
\let\crnotice\mycrnotice%
\let\confname\myconfname%

\permission{Permission to make digital or hard copies of all or part of this work for personal or classroom use is granted without fee provided that copies are not made or distributed for profit or commercial advantage and that copies bear this notice and the full citation on the first page. Copyrights for components of this work owned by others than the author(s) must be honored. Abstracting with credit is permitted. To copy otherwise, or republish, to post on servers or to redistribute to lists, requires prior specific permission and/or a fee. Request permissions from permissions@acm.org.}
\conferenceinfo{SPSM'13,}{November 8, 2013, Berlin, Germany. \\
{\mycrnotice{Copyright is held by the owner/author(s). Publication rights licensed to ACM.}}}
\copyrightetc{ACM \the\acmcopyr}
\crdata{978-1-4503-2491-5/13/11\ ...\$15.00.\\
http://dx.doi.org/10.1145/2516760.2516769}

\newcommand{\mylongtitle}{Sound and Precise Malware Analysis for Android via Pushdown Reachability and Entry-Point Saturation}

\newcommand{\mytitle}{\mylongtitle}

\newcommand{\ie}{\emph{i.e.}}

\newcommand{\etal}{\emph{et al.}}

\newcommand{\syn}[1]{\mathsf{#1}}

\newcommand{\s}[1]{\mathit{#1}}

\newcommand{\parto}{\rightharpoonup}

\newcommand{\setbuild}[2]{\left\{ #1 : #2\right\}}

\newcommand{\Pow}[1]{{\mathcal{P}\left(#1\right)}}
\newcommand{\PowSm}[1]{{\mathcal{P}(#1)}}

\newcommand{\vect}[1]{\langle #1\rangle}

\newcommand{\join}{\sqcup}

\newcommand{\bigjoin}{\bigsqcup}

\newcommand{\sembr}[1]{\ensuremath{[\![{#1}]\!]}}

\newcommand{\opor}{\mathrel{|}}

\newcommand{\produces}{\mathrel{::=}}

\newcommand{\stmt}{s}

\newcommand{\aexpr}{\mbox{\sl {\ae}}}

\newcommand{\store}{\sigma}

\newcommand{\addr}{a}

\newcommand{\aTo}{\leadsto}

\newcommand{\aInject}{{\hat{\mathcal{I}}}}

\newcommand{\sa}[1]{\widehat{\mathit{#1}}}

\newcommand{\aEval}{{\hat{\mathcal{E}}}}
\newcommand{\aArgEval}{{\hat{\mathcal{A}}}}

\newcommand{\aStore}{\sa{Store}}

\newcommand{\astore}{{\hat{\sigma}}}

\newcommand{\acont}{{\hat{\kappa}}}

\newcommand{\aden}{{\hat{d}}}

\newcommand{\aaddr}{{\hat{\addr}}}

\newcommand{\cexpr}{\mathit{ce}}

\newcommand{\afp}{\hat{\mathit{fp}}}

\newcommand{\reg}{\mathit{name}}
\newcommand{\aconf}{{\hat c}}

\newcommand{\aphrame}{\hat{\phi}}

\newcommand{\acallframe}{\widehat{\chi}}

\newcommand{\aexnframe}{\widehat{\eta}}

\newcommand{\aCallFrame}{\widehat{\mathit{CallFrame}}}

\newcommand{\aHandlerFrame}{\widehat{\mathit{HandlerFrame}}}

\newcommand{\aFieldAddr}{\sa{FieldAddr}}

\newcommand{\afa}{\sa{fa}}

\newcommand{\aRegAddr}{\sa{RegAddr}}

\newcommand{\ara}{\sa{ra}}

\newcommand{\aobjp}{\sa{op}}

\newcommand{\objv}{\mathit{ov}}
\newcommand{\aobjv}{\hat{\objv}}

\newcommand{\ObjectPointer}{\s{ObjectPointer}}
\newcommand{\aObjectPointer}{\widehat{\ObjectPointer}}

\newcommand{\FramePointer}{\s{FramePointer}}
\newcommand{\aFramePointer}{\widehat{\FramePointer}}

\newcommand{\FieldEval}{\mathcal{A}_{\mathcal{F}}}
\newcommand{\aFieldEval}{\hat\FieldEval}

\newcommand{\allocFP}{\s{allocFP}}
\newcommand{\allocOP}{\s{allocOP}}

\newcommand{\aallocFP}{\widehat{\allocFP}}
\newcommand{\aallocOP}{\widehat{\allocOP}}

\newcommand{\StmtOf}{{\mathcal{S}}}

\mathchardef\mhyphen="2D

\renewcommand{\footnoterule}{\hrule width .5\columnwidth height .51pt depth 0pt}
\makeatletter

\definecolor{dkgreen}{rgb}{0,0.6,0}
\definecolor{gray}{rgb}{0.5,0.5,0.5}
\definecolor{mauve}{rgb}{0.58,0,0.82}

\begin{document}


\title{\mytitle}


\author{
\alignauthor
Shuying Liang, Andrew W. Keep, Matthew Might,\\
Steven Lyde, Thomas Gilray, and Petey Aldous\\
\affaddr{University of Utah}\\
\email{\{liangsy,akeep,might,lyde,tgilray,petey.aldous\}@cs.utah.edu}
\and
\alignauthor
David Van Horn\\
\affaddr{Northeastern University}\\
\email{dvanhorn@ccs.neu.edu}
}

\maketitle

\begin{abstract}
Sound malware analysis of Android applications is challenging.
First, object-oriented programs exhibit highly interprocedural, dynamically dispatched control structure.
Second, the Android programming paradigm relies heavily on
the  asynchronous execution of multiple entry points.
Existing analysis techniques focus more on the second challenge,
while relying on traditional analytic techniques that suffer from inherent imprecision or unsoundness
  to solve the first.

We present Anadroid, a static malware analysis framework for Android apps.
Anadroid exploits two techniques to soundly raise precision:
(1) it uses a pushdown system to precisely model dynamically dispatched interprocedural \emph{and} exception-driven
    control-flow;
(2) it uses \textit{Entry-Point Saturation} (EPS) to soundly approximate all possible interleavings
 of asynchronous entry points in Android applications.
(It also integrates static taint-flow analysis and least permissions analysis to expand the class of malicious behaviors which it can catch.)
Anadroid provides rich user interface support  
for
human analysts which must ultimately rule on the ``maliciousness'' of a behavior.
%

To demonstrate the effectiveness of Anadroid's malware analysis, 
we had teams of analysts 
analyze
a challenge suite of 52 Android applications released as part of the
Automated Program Analysis for Cybersecurity (APAC) DARPA program.
The first team analyzed the apps using a version of Anadroid 
that uses traditional (finite-state-machine-based) control-flow-analysis
found in existing malware analysis tools;
the second team analyzed the apps using a version of Anadroid
that uses our enhanced pushdown-based control-flow-analysis.
We measured machine analysis time, 
human analyst time,
and their accuracy in flagging malicious applications.
With pushdown analysis, we found statistically significant ($p < 0.05$) decreases in time:
from 85 minutes per app to 35 minutes per app in human plus machine analysis time;
and statistically significant ($p < 0.05$) increases in accuracy
with the pushdown-driven analyzer: from 71\% correct identification to 95\% correct identification.
\end{abstract}




\section{Introduction}

Google's Android operating system is the most popular mobile platform, with a 52.5\% share of all smartphones~\cite{local:Gartner:url}.
Due to Android's open application development community, more than 400,000 apps are available 
with 10 billion cumulative downloads by the end of 2011~\cite{local:google10b:url}.

While most of these third-party apps have legitimate reasons to access private data, utilize the Internet, or make changes to local settings and file storage,
the permissions provided by Android are too coarse, allowing malware to slip through the cracks.
For instance, an app that needs to read information from only a specific website and access to GPS information
must, necessarily, be granted full read/write access to the entire Internet, allowing it to maliciously leak location information.
In another example, a note-taking application that writes notes to the file system can use the file system permissions to wipe out SD card files when a trigger condition is met.
Meanwhile, a task manager that legitimately requires every permission available can be 
benign.
To understand application behaviors like these before running the program, we need to statically analyze the application,
tracking what data is accessed, where sensitive data flows, and what operations are performed 
with the data, \ie, determine whether data is tampered with. 

However, static malware analysis for Android apps is challenging.
First, there is the general challenge of analyzing object-oriented programs where the state of the art is finite state based.
 Specifically,  traditional analysis regimes like
 $k$-CFA~\cite{mattmight:Shivers:1991:CFA} and its many variants implicitly or
 explicitly finitize the stack during abstraction.
 In effect, analyzers carve up dynamic return points and exception-handling
 points among a finite number of abstract return contexts.
 When two dynamic return points map to the same abstract context, the analyzer loses
 the ability to distinguish them.
 
Second, there is the domain-specific challenge from the Android programming paradigm, 
which is event-driven with multiple entry points and asynchronous interleaved execution. 
A typical strategy of existing static malware detection usually  relies upon 
an existing analytic framework, \ie, CHEX~\cite{shuyingliang:Lu:2012:CCS:CHEX} depends on WALA~\cite{local:wala:url}, Woodpecker depends on Soot~\cite{local:soot:url}, 
and various \emph{ad hoc} techniques proposed to deal with the problem: 
either by heuristic aborting of the paths unsoundly~\cite{shuyingliang:Lu:2012:CCS:CHEX}, 
or combining dynamic execution to eliminate  paths not appearing at run time~\cite{shuyingliang:Zheng:2012:SPSM:SmartDroid}.
One problem with this strategy is the inherent imprecision of the underlying analytic framework means imprecise  malware analysis on top of it.  
Another problem is that many static malware analyzers go unsound in order to handle the large number of permutations for multiple entry points for the sake of efficiency.
This means the static analyzer can no longer prove the absence of behaviors, malicious or otherwise.

In this paper, we describe Anadroid, a generic static malware analyzer for Android apps.
Anadroid depends on two integrated techniques:
(1) it is  the first  static analysis using higher-order pushdown
control-flow-analysis    
for an object-oriented programming language;  
(2)  it uses \textit{Entry-Point Saturation} (EPS) to soundly approximate interleaving execution of asynchronous entry points in Android apps.
We also integrate static taint flow analysis and least permissions analysis as application security analysis. 
In addition, Anadroid provides a rich user interface to support  
human-in-the-loop malware analysis.
It allows user-supplied predicates to filter and highlight analysis results.

To demonstrate the effectiveness of Anadroid's malware analysis, 
we evaluate a challenge suite of 52 Android apps released as part of the DARPA
Automated Program Analysis for Cybersecurity (APAC) program.
We compare 
a malware analyzer driven by finite-state control-flow-analysis
with  our model  with respect to  accuracy and analysis time. 
We  found that  
pushdown-driven malware analysis leads to statistically significant improvements in both aspects 
over traditional static analysis methods (which use finite-state methods like $k$-CFA to handle program control~\cite{mattmight:Shivers:1991:CFA}).

The remainder of this paper is organized as follows.
We illustrate the challenges via a condensed example malicious application in Section~\ref{sec:motivation-exam}.
Then we present our solutions in the subsequent two sections, 
where Section~\ref{sec:pushdown} presents the foundational pushdown control-flow analysis
and Section~\ref{sec:multi-entries}
details \textit{Entry-point saturation} (EPS).  
Based on pushdown control-flow analysis and EPS, we integrate static taint flow analysis and least permissions analysis into our analytic framework, as presented in Section~\ref{sec:other-techs}.
Section~\ref{sec:tool} presents our tool and Section~\ref{sec:eval} evaluates the effectiveness of our technique by comparing our results with results from finite state-based analysis. 
Case studies summarizing the vulnerabilities we have found are described in Section~\ref{sec:cases}.
Section~\ref{sec:related} presents related work and Section~\ref{sec:concl} concludes.

\section{Challenges of Static Malware Analysis for Android}\label{sec:motivation-exam}

The program \textit{Kitty.java} is adapted from an Android malware
application developed and released by a red team on the DARPA APAC program.
It exfiltrates location data from pictures stored on the users phone to a
malicious site (on Lines 19 and 33) or posts the information (on Lines 36--38)
via an Android Intent if the first attempt fails. 
This example helps illustrate the two challenges of static malware detection
for Android programs.

\begin{lstlisting}[language=Java]
public class KittyQuote extends Activity {
  String img = "k155"; // kitty image
  // other fields ...
  public void onCreate(Bundle savedState) {
    super.onCreate(savedState);
    // build quote list and other initialization
  }
  public String getKitty() {
    // access "DCIM/Camera" and use ExifInterface
    //  to exfiltrate location
  } 
  public void aboutButton(View view) { 
    // display normal information
    String website = "http://www.catquotes.com"; 
    startActivity(
      new Intent(Intent.ACTION_VIEW,
            Uri.parse(website)));
    try { 
      new SendOut().execute(website);
    } catch (Exception e) { }
  }
  public void nextButton(View view) { 
    img = getKittey();  // store loc info
  }
  public void prevButton(View view) {
    img = getKittey();
  }
  public void kittyQuoteButton(View view) {
    // display kitty quote as toast message 
    // ... and send out location info to a website
    String url = "http://www.catquotes.com?" + img; 
    try { 
      new SendOut().execute(url); 
    } catch (Exception e) { 
      // if network fails, not giving up 
      startActivity(
        new Intent(Intent.ACTION_VIEW,
              Uri.parse(url)));
    }
  } 
  class SendOut extends AsyncTask {
    protected Void doInBackground(URI... uris) {
      HttpGet hg = new HttpGet(uris[0]);
      try {
        new DefaultHttpClient().execute(hg);
      } catch (Exception e) { }
      return null;
    }
    // ... 
  } 
}
\end{lstlisting}

\paragraph{Fundamental challenge: imprecision induced by finite-based analytic
model for object-oriented programs}

Android apps are written in Java, and it can be difficult to statically
produce a precise control flow graph of the program, particularly
in the presence of exceptions.
Many existing static analyzers for Java programs are based on
finite-state-machine analysis, \ie~$k$-CFA or $k$-object sensitivity,
which have limited analytic power for analyzing dynamic dispatch or exceptions precisely.
Analyzers built using these techniques have a greater rate of false positives
and false negatives due to this imprecision.

For instance, in the example, both \textit{aboutButton} and
\textit{kittyQuoteButton} create a non-blocking background thread via the
class \textit{SendOut} (Lines 19 and 33) wrapped inside a try block.
In the \textit{aboutButton} code, normal exception handling occurs at Line 20.
In the \textit{kittyQuoteButton} code, the exception handler attempts a second
malicious action in Lines 36--38 using an Android web view activity to send out
location information.

This illustrates one of the difficulties in making a precise analysis of the
program, where the flow of the exception is difficult to track, leading to
false positives or false negatives.
Specifically, a finite-state-based analyzer cannot determine which exception
handler block will be used when an exception is thrown at Line 19.
It can (incorrectly) conclude that the handler block in Lines 36--38 will be
invoked.
This causes a false positive where Line 19 is identified as potentially leading
to the malicious behavior in Lines 36--38.
In the same way, a finite-state-based analyzer cannot determine which exception
handler block will be used when an exception is thrown from Line 33.
It may determine that the handler block on Line 19 is the appropriate exception
handler block.
This leads to a false negative, where the malicious behavior resulting from an
exception at Line 33 is missed.
The pushdown control-flow analysis model can precisely track exception handling, so our analyzer
identifies the correct handler block (Line 20) where no malicious behavior
exists.

In fact, the fundamental imprecision caused by spurious control flows due to
exceptions has been reported before.
Fu~\etal{}~\cite{Fu:2005:rubust-java-server-apps} report this problem when testing Java server
apps, and
Bravenboer~\etal{}~\cite{Bravenboer:2009:Exceptions,Bravebboer:2009:declare-pointsto}
describe the need to combine points-to and exception analysis to attempt to
regain some of this precision.
Unlike approaches using finite-state-based control flow analysis, the pushdown
control-flow analysis
used as the foundation for our malware analyzer can precisely match both normal
and exception return flows achieving lower false positive and false negative
rates.

\paragraph{Domain specific challenge: Permutations of Asynchronous multi-entry-points}
 
The second challenge in analyzing Android apps is caused by the
asynchronous multiple entry points into an Android application.
The Android framework allows developers to create rich, responsive, and
powerful apps by requiring developers to organize their code into
components.
Each component type serves a different purpose:
(1) activities for the main user-interface,
(2) services for non-blocking code or remote processes,
(3) content providers for managing application data, and
(4) broadcast receivers to provide system-wide announcements.
Applications can register various component handlers, either explicitly in code
or through a resource file (res/layout/\textit{filename}.xml).
Whenever an event occurs, the callbacks for the event are invoked
asynchronously, potentially interleaving their execution with those in other components.
Different apps can also invoke each other by exposing functionality via
an \textit{Intent} at both the application and component level\footnote{Apps sharing the same Linux user ID are also able to access each other's
files.}.
Unlike an application with a single entry point, static analysis for an Android
application must explore all permutations of these asynchronous entry points.
Analyzing all permutations can greatly increase the expense of the analysis.
As a result many analyzers use an unsound approximation that can lead to false
negatives.

Our example illustrates these problems.
First, not all of the callback methods are explicitly registered in code, like the
  \textit{onCreate} method of \textit{KittyQuote} and the \textit{doInBackground}
 method of \textit{SendOut}, \textit{aboutButton}, \textit{nextButton},
\textit{prevButton}, and \textit{kittyQuoteButton}\footnote{In code, these are 
implemented with the \textit{onClick} method in an
\textit{OnClickListener}.} are registered in an XML layout resource file as follows.
 
\begin{lstlisting}[language=XML, frame=none, numbers=none, captionpos=none]
<Button
        android:id="@+id/button2"
        ...
        android:onClick="prevButton" />
\end{lstlisting}

Second,  malicious behaviors can be
triggered from an entry point or series of entry points.
For instance, if the \textit{prevButton} or \textit{nextButton} methods are
called before the \textit{aboutButton} or \textit{kitteyQuoteButton} methods,
the application will leak location data gathered from the Exif data of pictures
on the device.
In order to avoid missing malicious behavior, while still performing the
analysis efficiently, we need a way to approximate the possible permutations of
the asynchronous multi-entry points without loosing soundness.
This means we cannot use heuristic pruning, but we also do not want to use a
dynamic analysis, since we hope to analyze the program before we attempt to run
it.
This leads to our second contribution, Entry-Point Saturation (EPS),
which can be used to analyze multi-entry points in a sound way.

To summarize the relationship between the challenges, an unsound approximation
of asynchronous entry points can miss malicious behavior,
while any imprecision in the underlying control-flow analysis can result 
in an analysis that misses malicious behavior in programs.
This problem is further exacerbated by additional highly dynamic dispatched inter-procedural control flows 
 caused by permutations of  entry points. 
We need to address both challenges to ensure our analyzer does not miss
malicious behavior in the program.
The next two sections describe our solutions to these challenges.

\section{Pushdown flow analysis for OO}\label{sec:pushdown}

In this section, we describe our pushdown control-flow-analysis for
object-oriented programs.
We present the analysis as a small-step semantic analysis following the style
of Felleisen~\etal{}~\cite{mattmight:Felleisen:1987:CESK} and more recently
Van Horn~\etal{}~\cite{mattmight:VanHorn:2010:Abstract}. 
We first define an object-oriented bytecode language closely modeled on 
Dalvik bytecode\footnote{Java code in Android apps are compiled
to Dalvik bytecode.}.
Our language dispenses with some of the bytecode size optimizations that the
Dalvik bytecode uses to shrink the size of programs, allowing the analyzer to
treat Dalvik instructions with similar functionality as a single instruction.
Our language also includes explicit line number instructions to allow it to
more easily relate malicious behavior in the bytecode to the original source
code.
Then we develop our pushdown control-flow analysis for this
language\footnote{Although Android apps can include native code, our analyzer
only handles Dalvik bytecode.  Native code in the Android API is modeled
directly in the analyzer.}.

\subsection{Syntax}
 
The syntax of the bytecode language is given in Figure~\ref{fig:oo-syntax}.
Statements encode individual actions for the machine;
Complex expressions encode expressions with possible 
side effects or non-termination; and
atomic expressions encode atomically computable values.
There are four kinds of names:
$\syn{Reg}$ for registers, $\syn{ClassName}$ for class names,
$\syn{FieldName}$ for field names,
and $\syn{MethodName}$ for method names.
There are two special register names:
$\syn{ret}$, which holds the return value of the last function called,
and $\syn{exn}$, which holds the most recently thrown exception.

The syntax is a straightforward
abstraction of Dalvik bytecode,
but it is worth examining statements related to exceptions 
in more detail:

 \begin{itemize}
 \item $(\syn{throws}~\mathit{class}\mhyphen\mathit{name} \dots)$
 indicates that a  method can throw one of the named exception classes,
\item $(\syn{push}\mhyphen\syn{handler}~\mathit{class}\mhyphen\mathit{name}~\mathit{label})$
pushes an exception handler frame on the stack that will catch 
 exceptions of type $\mathit{class}$ 
 and divert execution to $\mathit{label}$, and
 
 \item
 $(\syn{pop}\mhyphen\syn{handler})$
 pops the initial exception handler frame off the stack.

 \end{itemize}

\begin{figure}
{\fontsize{8.5pt}{8.5pt}\selectfont
 \begin{align*}
\mathit{program} &\produces 
\mathit{class} \mhyphen \mathit{def}\; \ldots
\\
\mathit{class} \mhyphen \mathit{def} &\produces (\mathit{attribute}\;\ldots~ \syn{class} ~\mathit{class}\mhyphen\mathit{name} ~\syn{extends}  ~\mathit{class}\mhyphen\mathit{name} 
\\
&\;\;\;\;\;\;\;\; (\mathit{field}\mhyphen{\mathit{def}}  \dots)\; (\mathit{method}\mhyphen\mathit{def} \dots))
\\
\mathit{field}\mhyphen\mathit{def} &\produces  (\syn{field}~\mathit{attribute}\;\ldots~\mathit{field}\mhyphen\mathit{name}~\mathit{type})
\\
\mathit{method}\mhyphen\mathit{def} &\produces  (\syn{method}~\mathit{attribute}\;\ldots~\mathit{method}\mhyphen\mathit{name}~(\mathit{type} \dots)~\mathit{type} 
\\
&\;\;\;\;\;\;\;\;
(\syn{throws}~\mathit{class}\mhyphen\mathit{name} \dots)
\;
(\syn{limit} ~n)
\; 
\stmt\; \ldots)
\\
\stmt \in \syn{Stmt}&\produces (\syn{label} ~\mathit{label}) \opor (\syn{nop}) \opor (\syn{line} ~\mathit{int}) \opor  (\syn{goto} ~\mathit{label}) 
\\
&\;\;\opor\;\;  (\syn{if}\; \aexpr ~(\syn{goto}~\mathit{label})) \opor (\syn{assign}~\mathit{\reg}\; [\aexpr \opor \cexpr]) 
\\
&\;\;\opor\;\; (\syn{field}\mhyphen\syn{put} ~\aexpr_o ~\mathit{field}\mhyphen\mathit{name} ~\aexpr_v)  
\\
&\;\;\opor\;\;  (\syn{field}\mhyphen\syn{get}\; \reg ~\aexpr_o ~\mathit{field}\mhyphen\mathit{name})
\\
&\;\;\opor\;\; (\syn{push}\mhyphen\syn{handler} ~\mathit{class}\mhyphen\mathit{name} ~\mathit{label}) 
\\
&\;\;\opor\;\; (\syn{pop}\mhyphen \syn{handler}) \opor (\syn{throw}~\aexpr) \opor (\syn{return} ~\aexpr) 
\\
\aexpr \in \syn{AExp} &\produces \syn{this} \opor \syn{true} \opor \syn{false} \opor \syn{null} \opor \syn{void} 
\\
&\;\;\opor\;\; \mathit{\reg} \opor \mathit{int} 
\\
&\;\;\opor\;\;  (\mathit{atomic}\mhyphen\mathit{op} ~\aexpr \dots \aexpr)  
\\
 &\;\;\opor\;\; \syn{instance}\mhyphen\syn{of}(\aexpr, \mathit{class}\mhyphen\mathit{name}) 
\\
\cexpr &\produces (\syn{new}\; \mathit{class}\mhyphen\mathit{name}) 
\\
&\;\;\opor\;\;(\mathit{invoke}\mhyphen\mathit{kind} ~(\aexpr\dots\aexpr) (\mathit{type}\;\dots\; ))
\\
\mathit{invoke}\mhyphen\mathit{kind} &\produces \syn{invoke}\mhyphen\syn{static}  \opor \syn{invoke}\mhyphen\syn{direct}   \\
&\;\;\opor\;\;\syn{invoke}\mhyphen\syn{virtual}  \opor \syn{invoke}\mhyphen\syn{interafce}    \\
&\;\;\opor\;\;\syn{invoke}\mhyphen\syn{super}
\\
\mathit{type} &\produces ~\mathit{class}\mhyphen\mathit{name} \opor \syn{int} \opor \syn{byte} \opor \syn{char} \opor \syn{boolean}
\\
\mathit{attribute} &\produces \syn{public} \opor \syn{private} \opor \syn{protected} 
\\
&\;\;\opor\;\; \syn{final} \opor \syn{abstract} \\
\reg & ~\text{is an infinite set of frame-local variables} 
\\
&\text{or registers in the parlance of Dalvik byte code}
\text.
\end{align*}
\caption{An object-oriented bytecode adapted from the Android specification~\cite{local:androidbytecode:url}.}

\label{fig:oo-syntax}
}
 
\end{figure}
 
 With respect to a given program, we assume a syntactic metafunction
 $\StmtOf : \syn{Label} \to \syn{Stmt}^*$, which maps a label to the
 sequence of statements that start with that label.

\subsection{Abstract Semantics}
With the language in place, the next step is to define the concrete semantics of the language, 
providing the most accurate interpretation of program behaviors.
However, the concrete semantics cannot be used directly for static analysis, since it is not computable.
Fortunately, we can adapt the technique of abstracting abstract
machines~\cite{mattmight:VanHorn:2010:Abstract} to derive a sound abstract
semantics for the Dalvik bytecode using an abstract CESK machine.

  States of this machine consist of a series
     of statements, a frame pointer, a heap, and a stack.  
 The
   evaluation of a program is defined as the set of \emph{abstract}
   machine configurations reachable by an abstraction of the machine
   transition relation. Largely, abstract evaluation is defined as $\aEval :
   \syn{Stmt^*} \to \PowSm{\sa{Conf}}$, where   $\aEval(\vec{s}) = \setbuild{ \aconf }{ \aInject{(\vec{s}) \aTo^* \aconf }}. $
   Therefore, abstract evaluation is defined by the set of configurations reached by
   the reflexive, transitive closure of the $(\aTo)$ relation, which shall be defined in Section~\ref{sec:abs-transition-rules}.

\paragraph{Abstract configuration-space} Figure~\ref{fig:abs-conf-space} details the abstract configuration-space for this abstract machine. 
We assume the natural element-wise, point-wise, and member-wise
lifting of a partial order across this state-space.
 \begin{figure}
\begin{center}
{\fontsize{8.5pt}{8.5pt}\selectfont
\begin{align*}
 \aconf \in \sa{Conf} &= \syn{Stmt^*} \times \sa{FramePointer} \times \sa{Store} \times \sa{Kont} 
\\
 \astore \in \sa{Store} &= \sa{Addr}  \parto   \sa{Val} 
 \\
\aaddr \in  \sa{Addr} &=  \aRegAddr  + \aFieldAddr 
 \\
  \ara \in \aRegAddr &= \sa{FramePointer}  \times \syn{Reg}  
  \\
 \afa \in \aFieldAddr &= \sa{ObjectPointer} \times {\syn{FieldName} }
\\
\acont \in \sa{Kont} &= \sa{Frame}^* 
\\
\aphrame \in \sa{Frame} &= \aCallFrame + \aHandlerFrame    
\\
\acallframe \in \aCallFrame &\produces \mathbf{fun}(\afp,\vec{s})
\\
\aexnframe \in \aHandlerFrame &\produces \mathbf{handle}(\mathit{class\text{-}name},\mathit{label})
\\
\aden \in \sa{Val} &= \Pow{ \sa{ObjectValue} +  \sa{String} +  \sa{\mathcal{Z}}} 
 \\
 \aobjv \in \sa{ObjectValue} &= \aObjectPointer \times \syn{ClassName}
\\
\afp \in \aFramePointer &\text{ is a finite set of frame pointers} 
\\ 
\aobjp \in \aObjectPointer&\text{ is a finite set of object pointers} 
\text.
 \end{align*}
\caption{The abstract configuration-space.}
\label{fig:abs-conf-space}
}
\end{center}
 \end{figure}

    
    To synthesize the abstract state-space,
    we force frame pointers and object pointers (and thus addresses) to be a finite set.
    When we restrict the set of addresses to a finite set, 
    the machine may run out of addresses to allocate, and when it does, 
    the pigeon-hole principle will force multiple abstract values to reside at the same address. 
    As a result, the range of the $\aStore$ becomes a power set 
    in the abstract configuration-space. 
   Crucially, in this machine, the stack is left unbounded, unlike the final step in the abstracting abstract machine approach. 
   This enables us to faithfully model both normal function call and return and exception throw and catch handling intraprocedurally and interprocedurally.
    %
   In the next section we detail  the essence of the abstract CESK machine, which is one of the main contributions of this work.
    
%

\subsubsection{Abstract transition relation\label{sec:abs-transition-rules}}

The machine relies on helper functions to evaluate atomic
expressions and look up field values:

\begin{itemize}

\item
$\aInject : \syn{Stmt^*} \to \sa{Conf}$ injects a sequence of instructions into
                                              a configuration:
 \begin{equation*}
\aconf_0 = \aInject(\vec{s}) = (\vec{s}, \afp_0, [], \vect{})
  \text.
  \end{equation*}

\item
$\aArgEval : \syn{AExp} \times \aFramePointer \times \aStore \rightharpoonup
  \sa{Val}$ evaluates atomic expressions:
  \begin{align*}
    \aArgEval(\reg,\afp,\astore) &= \store(\afp,\reg) && \text{[variable look-up]}
    \text.
    \end{align*}

\item
$ \aFieldEval : \syn{AExp} \times  \aFramePointer \times \aStore \times \syn{FieldName} \rightharpoonup \sa{Val}$
looks up fields:
  \begin{align*}
    \aFieldEval(\aexpr_o,\afp,\astore, \mathit{field}\mhyphen \mathit{name}) &=  \bigjoin \astore(\aobjp,\mathit{field}\mhyphen \mathit{name}) 
    \\ 
    ~\text{where} ~(\aobjp, \mathit{class}\mhyphen \mathit{name}) &\in  \aArgEval(\aexpr_o, \afp, \astore )
    \text.
    \end{align*}

\end{itemize}

 The abstract transition relation $(\aTo) \subseteq \sa{Conf} \times
    \sa{Conf}$ has rules to
soundly model all possible concrete executions of a bytecode program.
In the subsequent subsections, we illustrate the rules that involve objects,
function calls, and exceptions, omitting more obvious rules to save space.
 
\paragraph{New object creation}
Creating a new object allocates a potentially non-fresh address and joins the
newly initialized object to other values residing at this store address.

 $   \overbrace{(\sembr{(\syn{assign}\; \reg\; (\syn{new}\;\mathit{class}\mhyphen\mathit{name})) :  \vec{s}}, \afp, \astore,  \acont)}^{\aconf}
  $ 
    \begin{align*}
   &\aTo   (\vec{s}, \afp, \astore'',  \acont), ~\text{where}
      \\
   \aobjp' &= \aallocOP(\aconf)
   \\
   \astore' &= \astore \join [(\afp, \mathit{\reg}) \mapsto  ( \aobjp', \mathit{class}\mhyphen\mathit{name})]
   \\
  \astore'' &= \widehat{\mathit{initObject}}(\astore', \mathit{class}\mhyphen\mathit{name})
  \text{,}
    \end{align*}
The  helper function
$\widehat{\mathit{initObject}}: \aStore \times \syn{ClassName} \rightharpoonup \aStore$ initializes the object's fields. 
  
  \paragraph{\textbf{Instance field reference/update}}
Referencing a field uses $\aFieldEval$ to lookup the field values and joins these values with the values at the store location for the destination register:

 $ (\sembr{(\syn{field}\mhyphen\syn{get}\; \reg ~\aexpr_o ~\mathit{field}\mhyphen\mathit{name}) : \vec{s}}, \afp, \astore,\acont)$
 \begin{align*}
   &\aTo
     (\vec{s},\afp,\astore', \acont)
  \text{, where}
  \\
  \astore' &= \astore \join [ (\afp, \reg)\mapsto \aFieldEval(\aexpr_o, \afp, \astore, \mathit{field}\mhyphen\mathit{name} ) ] 
  \text.
   \end{align*}
   Updating a field first determines the abstract object values from the store,
   extracts the object pointer from all the possible values,
   then pairs the object pointers with the field name to get the field address,
   and finally \textit{joins} the new values to those found at this store location:
   
  $   (\sembr{(\syn{field}\mhyphen\syn{put}~\aexpr_o~\mathit{field}\mhyphen{name} ~\aexpr_v) : \vec{s}}, \afp, \astore, \acont)
  $
   \begin{align*}
      &\aTo(\vec{s},\afp,\astore', \acont)~~\text{where}\\
  \astore' &= \astore \join [(\aobjp, \mathit{field}\mhyphen{name}) \mapsto \aArgEval(\aexpr_v, \afp, \astore)] 
  \\
  (\aobjp, &\mathit{class}\mhyphen{name}) \in \aArgEval(\aexpr_o, \afp, \astore)\text.
   \end{align*}

\paragraph{Method invocation}
This rule involves all four components of the machine.
The abstract interpretation of non-static method invocation can result in the
method being invoked on a \emph{set} of possible objects, rather than a single
object as in the concrete evaluation.
Since multiple objects are involved, this can result in different method definitions being resolved for the different objects.
The method is resolved\footnote{Since the language supports inheritance, method
resolution requires a traversal of the class hierarchy.  This traversal follows
the expected method and is omitted here so we can focus  on the abstract
rules.} and then applied as follows: 

 \begin{align*}
  \overbrace{
    (\sembr{ (\mathit{invoke}\mhyphen\mathit{kind} ~(\aexpr_0\dots\aexpr_n)\; (\mathit{type}_0\dots\mathit{type}_n))} :  \vec{s}, \afp, \astore, \acont)
    }^{\aconf}
    \\
  \aTo 
     \widehat{\mathit{applyMethod}}(m, \vec{\aexpr}, 
     \afp,\astore,\acont)
\text,
  \end{align*}
where the  function ${\widehat{\mathit{applyMethod}}}$ takes a method definition,
arguments,
a frame pointer,
a store, and a new continuation  and produces the next configuration:

$ \widehat{\mathit{applyMethod}}(m, \vec{\aexpr}, \afp, \astore, \acont) = (\vec{\stmt}, \afp', \astore', (\afp, \vec{s}) : \acont),$ where
\begin{align*}
\afp' &=  \aallocFP(\aconf)
\\
 \astore' &= \astore \join [(\afp', \reg_i) \mapsto \aArgEval(\aexpr_i, \afp, \astore)]\text.
\end{align*}

\paragraph{\textbf{Procedure return}}
Procedure return restores the caller's context and \textit{extends} the return value in
the dedicated return register,  $\syn{ret}$.

$(\sembr{(\syn{return}~\aexpr) :  \vec{s}}, \afp, \astore,  \mathbf{fun}(\afp',\vec{s}') : \acont)
    \aTo 
       (\vec{s}', \afp', \astore',  \acont)\text, $%

$
   \text{where}~~
   \astore' = \astore\join [(\afp', \syn{ret})\mapsto \aArgEval(\aexpr, \afp, \astore)]
   \text.
$

If the top frame is an exception handler ($\mathbf{handle}$) frame, 
the abstract interpreter pops until the top-most frame is 
a function call ($\mathbf{fun}$) frame:
 \begin{align*}
      (\sembr{(\syn{return} ~\aexpr)} :  \vec{s}, \afp, \astore
      ,
      \mathbf{handle}(\mathit{class}\mhyphen\mathit{name}~\mathit{label}) : \acont)
  \\
  \aTo 
     (\sembr{(\syn{return} ~\aexpr)}:  \vec{s}, \afp,\astore
     , \acont)
 \text{.}
  \end{align*}

  \paragraph{Pushing and popping handlers}
Pushing and popping exception handlers is straightforward:
   \begin{align*}
      (\sembr{(\syn{push}\mhyphen\syn{handler} ~\mathit{class}\mhyphen\mathit{name}~\mathit{label})} :  \vec{s}, \afp, \astore, \acont)
    \\
    \aTo 
       (\vec{s},\afp,\astore,\mathbf{handle}(\mathit{class}\mhyphen\mathit{name}~\mathit{label}) : \acont)
    \end{align*}
   \begin{align*}
      (\sembr{(\syn{pop}\mhyphen\syn{handler})} :  \vec{s}, \afp, \astore, \mathbf{handle}(\mathit{class}\mhyphen\mathit{name}~\mathit{label}) : \acont)
    \\
    \aTo 
       (\vec{s},\afp,\astore, \acont)
   \text.
    \end{align*}
  \paragraph{Throwing and catching exceptions}
  The throw statement pops entries off the stack until it finds a matching exception handler:
   \begin{align*}
      (\sembr{(\syn{throw}~\aexpr)} :  \vec{s}, \afp, \astore
      ,\acont)
    &\aTo
    \widehat{\mathit{handle}}(\aexpr, \vec{s}, \afp, \astore, \acont)
   \text,
    \end{align*}
  where the function 
  $\widehat{\mathit{handle}} : \syn{AExp} \times \syn{Stmt^*} \times  \aFramePointer  \times \aStore \times \sa{Kont}$
  $\rightharpoonup \sa{Conf}$
  behaves like its concrete counterpart when the top-most frame is a
  compatible handler:
\begin{align*}
   &\widehat{\mathit{handle}}(\aexpr, \vec{s}, \afp, \astore,  \mathbf{handle}(\mathit{class}\mhyphen\mathit{name'}~\mathit{label}) : \acont')
   \\
   &\;\;\;\;= 
     (\StmtOf(\mathit{label}), \afp, \astore \join [(\afp, \syn{exn}) \mapsto( \aobjp, \mathit{class}\mhyphen\mathit{name})], \acont') 
   \text.
\end{align*}
Otherwise, it pops a frame:
\begin{align*}
    \widehat{\mathit{handle}}(\aexpr, \vec{s}, \afp, \astore,  \mathbf{handle}(\_,\_) : \acont')
    \\
    =
     (\sembr{(\syn{throw}~\aexpr)} :  \vec{s}, \afp, \astore
   , \acont') 
\\
    \widehat{\mathit{handle}}(\aexpr, \vec{s}, \afp, \astore,  \mathbf{fun}(\_,\_) : \acont')
    \\
    =
     (\sembr{(\syn{throw}~\aexpr)} :  \vec{s}, \afp, \astore
   , \acont') 
    \text.
\end{align*}

Executing the analysis consists of solving for the 
reachable control states of the implicit pushdown system
in the abstract semantics.
To compute the reachable 
control states of this pushdown system,
we
employ standard reachability algorithms from Reps'~\etal{}~\cite{mattmight:Bouajjani:1997:PDA-Reachability,mattmight:Kodumal:2004:CFL,mattmight:Reps:1998:CFL,mattmight:Reps:2005:Weighted-PDA}
work on pushdown-reachability.

\section{Entry-Point Saturation (EPS)}\label{sec:multi-entries}

The pushdown control-flow analysis described in the previous section provides the foundation for our object-oriented analysis.
Now, we shift our focus to addressing the domain specific challenge: asynchronous multiple entry points using Entry-Point Saturation (EPS) and integrating it into 
pushdown control-flow analysis.

An entry point is defined as any  point through which the system can enter the user application~\cite{local:android-funda:url}.
This means that any method that can be invoked by the framework is an entry point.
Since there is no single ``main'' method, the static analysis must first identify the entry points in the program.
Entry-point discovery is not a challenge, however, since they are defined by the Android framework.
We briefly summarize possible entry-points here.

There are three categories of entry points, which we generalize as \textit{unit}s. 
First, all the callback events of components defined by the Android framework are entry points.  
These  entry points are designed to be overridden by application code and are invoked and managed by the framework for the purpose of component life cycle management, coordinating among different components, and responding to user events which are themselves defined to be asynchronous. 
 Second, asynchronous operations that can be executed in the background by the framework are considered to be entry points. 
 These include the \textit{AsyncTask} class for short background operations, the \textit{Thread} class for longer operations, and the \textit{Handler} class for responding to messages.
Finally, all event handlers in Android UI widgets, such as button, check box, etc. are entry points.
Each UI widget has standard event listeners defined, where the event handler interface methods are meant to be implemented by application code.  
Entry points from the first two categories are found by parsing Dalvik bytecode and organized into a set attached to the corresponding \textit{unit}.
Entry points from the final category can also be defined in resource layout files \textit{res/layout/\textit{filename}.xml}, as illustrated in Section~\ref{sec:motivation-exam}.
These entry points can be obtained by parsing resource files before analysis.
 
\begin{figure}
  \centering
   \includegraphics[scale=0.5]{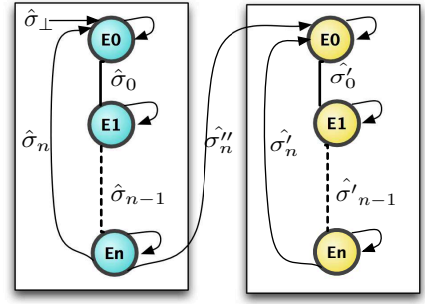}
     \caption{Entry-Point Saturation}
      \label{fig:mep}
      \end{figure}

After the entry point set for each unit is determined the real challenge begins.
If we want to do a sound analysis, all the permutations of the entry points need to be considered.
Complicating things further, entry points in the second category involve threads
so interleaved execution of these entry points is possible.
Other analyzers, such as CHEX, deal with this complexity using \textit{app-splitting}, which is unsound and cannot model the threading case. 
Here, we present a sound and efficient technique which directly relies on  the underlying pushdown analytic engine presented in Section~\ref{sec:pushdown}. 
Figure~\ref{fig:mep} illustrates the process.

 For each entry point $E_i$ in a \textit{unit} (represented as a square), 
 we compute the fixed point via pushdown analysis (refer to Section~\ref{sec:abs-transition-rules}).
After one round of computation the analysis returns a set of configurations. 
We then use the \textit{configuration widening} technique from Might~\cite{mattmight:Might:2007:Dissertation}  on the  set of configurations to generate a widened $\astore_i$.
This abstract component will be ``inherited'' by the next entry point in the fixed point computation.
The process repeats until the last entry point finishes its computation in a \textit{unit}. 
In this way, the unit has reached a fixed point.
The next step computes the fixed point between \textit{unit}s.
This is computed in a similar fashion to the intra-unit fixed point computation, 
so the widened result $\astore'_n$ from the  previous unit (the left square) 
participates in the reachability analysis of the next unit (the right square).

EPS soundly models all permutations of entry points and their interleaving
execution by passing the store resulting from analyzing one entry-point as the
initial store for the next entry-point.
This gives the ``saturated'' store for a given component.
The saturated store for a component is similarly passed as the initial store
for the next component.
The final store models every execution path without needing to enumerate every
combination.

EPS has several advantages:
\begin{enumerate}
\item It computes the fixed point from bottom up, intra-entry point,  inter-entry point, and inter-unit, in an efficient manner and
 significantly simplifies the static analysis; 
\item It not only computes all permutations of entry points, but also interleaving executions of them;
\item It is sound and easy to prove, because it is based on widening on the abstract configurations, 
where the soundness proof  follows the same 
structure as shown by Might~\cite{mattmight:Might:2007:Dissertation} and so we omit it here.
\end{enumerate}

Applying this technique to the example in Section~\ref{sec:motivation-exam}, we
can see that all execution sequences where location information is read
(\textit{nextButton} and \textit{prevButton}) to where it is leaked through
the ACTION\_VIEW Intent or the \textit{doInBackground} method of the
\textit{SendOut} AsyncTask are covered and analyzed.

\section{Taint flow analysis and Least\\ permissions analysis}\label{sec:other-techs}

As a whole, the previous two sections have described a solid foundation for static malware analysis. 
This section demonstrates two possible applications that build upon the foundation for specific security analyses.

\subsection{Pushdown taint flow analysis with entry-point saturation}
For any mobile application, one of the biggest concerns is leakage or tampering with private data.
We can easily instrument our framework to perform taint flow analysis to detect these malicious behaviors. 
The idea is adapted from early work by Liang~\etal{}~\cite{Liang:2012:HTA}, 
but has been enriched with Android sensitive data categories as taint values.
The taint flow analysis within our pushdown-based analysis framework 
is done by modifying the abstract configuration in Figure~\ref{fig:abs-conf-space} as:

{\fontsize{9.5pt}{9.5pt}\selectfont
$
\begin{array}{rcl}
 \sa{Conf}&  = & \syn{Stmt^*} \times \sa{FramePointer} \times \sa{Store} \times \\
          & &\sa{\textbf{\textit{TaintStore}}}\times \sa{Kont}
\end{array}
 $ }

and adding the definition for $\sa{\textbf{\textit{TaintStore}}}$, which is a flat lattice across all taint values:
{\fontsize{8.5pt}{8.5pt}\selectfont
\begin{align*}
 \astore^T \in \sa{\textbf{\textit{TaintStore}}} &=   \sa{Addr}  \parto   \sa{Val} 
 \\
 \aden \in \sa{Val} &= \Pow{ \sa{ObjectValue} +  \sa{String} +  \sa{\mathcal{Z}}  + \mathit{TaintVal}} 
 \\
 \mathit{TaintVal} &= \text{Location}  + \text{FileSystem} + \text{Sms} + \text{Phone}
 \\ &~~~~ + \text{Voice} +\text{DeviceID}+ \text{Network}+ \text{ID}
 \\&~~~~ + \text{TimeOrDate}
 +\text{Display} + \text{Reflection} 
 \\ &~~~~ + \text{IPC}+ \text{BrowserBookmark} + \text{SdCard}
 \\ &~~~~ +\text{BrowserHistory}+ \text{Thread}+ \text{Picture}
 \\
 &~~~~+ \text{Contact} +\text{Sensor} + \text{Account} + \text{Media}
 \text.
\end{align*}}

All the transition rules have an additional $\astore^T$ added, operations resemble the ones on $\astore$, 
except that the taint values are \textit{monotonically} propagated through the abstract semantics.
For example, in a function call, tainted values in arguments are bound to the formal parameters of the functions 
(using the $\join$ operation in the taint store $\astore^T$), returning abstract taint values bound to a register address with $\syn{ret}$, etc.
The detailed formalism is omitted to save space.

The taint propagation analysis is not limited to detecting sensitive data leakage or tampering.
As we discuss in Section~\ref{sec:cases},  it can help analysts to find  malicious behaviors, such as
 when SMS messages are blocked by a trigger condition or a limited resource is consumed, such as the local file system on the device being filled.
  
  \subsection{Least permissions analysis}
Our  analysis framework can also detect malicious behaviors in the presence of zero-permissions, as the motivating example demonstrates. 
However, for apps that request permissions, it is highly desirable 
to analyze how the apps use the permissions, 
since over-privileged apps can easily be exploited by other apps in the Android framework. 
It is straightforward to determine this situation in our framework by
instrumenting it with knowledge about permissions. 
Specifically, we use the data set from PScout~\cite{shuyingliang:Au:2012:CCS:PScout} to annotate each API call with permissions that are required for usage. 
During reachability analysis, permissions are inferred and collected. 
The reached permissions are determined during the analysis.
When the analysis finishes these permissions can be compared with the set of permissions requested in the manifest file.
This allows us to statically determine whether an application is over-privileged.

\section{The Tool: Anadroid} \label{sec:tool}

\begin{figure}
\centering
 \includegraphics[width=\columnwidth]{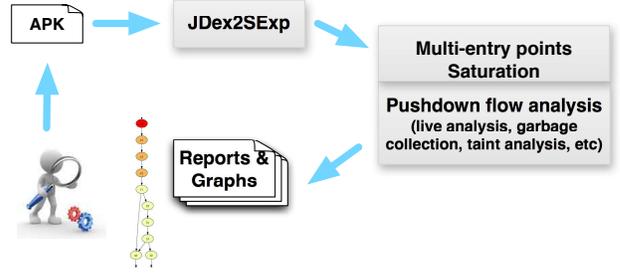}
   \caption{Software Architecture}
    \label{fig:anadroid-archi}
    \end{figure}

Anadroid is built on the principles illustrated in the previous sections.
Figure~\ref{fig:anadroid-archi} briefly sketches the software architecture.

Anadroid has the following features: 

\begin{itemize}
\item It consumes off-the-shelf Android application packages (files with suffix $\syn{.apk}$). In Figure~\ref{fig:anadroid-archi} \textit{JDexSExp} extracts the $\syn{.dex}$ file by invoking $\syn{apktool}$~\cite{local:apktool:url} and then disassembles binaries and generates an S-expression IR based on the $\syn{smali}$~\cite{local:Smali:url} format;
\item It enables human-in-the-loop analysis, by providing a rich user interface for an analyst to configure the analyzer, \ie setting the $k$ value, configuring abstract garbage collection to be used or not,  specifying predicates over the state space, etc., to trade off precision and performance, as well as semantic predicates to search analysis results;
\item It generates various reports and state graphs.
The following three reports are included:
(1) Least permissions presents which permissions are requested by an app and which permissions are inferred by Anadroid, reporting whether the app requests more permissions than it actually uses.
(2) The information flow report presents triggers (mainly UI triggers) and tainted paths that lead from sources to sinks,  with contexts such as class files, method names, and line numbers.
(3) The heat map report shows rough profiling results of the analyzer, which can be used to help an analyst understand where the analysis has focused its efforts and might indicate where an app developer has attempted to hide malicious behavior.
Analysis graphs are presented with an SVG formatted file as a reachable control flow graph.
It highlights suspicious source and sink states, as well as showing tainted paths between them.
In addition, an analyst can click on any state node in the graph for detailed inspection of the abstract execution at this point in the graph.
\end{itemize}

In fact, the pushdown analysis based analyzer  evolved from an older version with a similar user interface and output forms (reports and graphs).
The previous version of Anadroid used traditional finite-state methods (mainly $k$-CFA and $k$-object-sensitivity) that are employed in analysis platforms such as WALA~\cite{local:wala:url}, Soot~\cite{local:soot:url}, etc.

The old analyzer enables us to compare the traditional finite-state based
analysis with our new pushdown control-flow analysis to help us determine whether the new analysis is
more effective and efficient.
The following section details this comparison.

\section{Evaluation}\label{sec:eval}

We had two teams of analysts 
analyze
a challenge suite of 52 Android apps released as part of the DARPA
Automated Program Analysis for Cybersecurity (APAC) program.
Both teams are composed of the same five people, a mixture of graduate students
and undergraduate students, however, the applications are shuffled, ensuring
the same app is not evaluated by the same analyst.
The first team analyzed the apps with a version of Anadroid 
that uses traditional (finite-state-machine-based) control-flow-analysis
used in many existing malware analysis tools;
the second team analyzed the apps with a version of Anadroid
that uses our enhanced pushdown-based control-flow-analysis.
We measure the time the analyzer takes,
the time human analysts spent reviewing the results,
and the accuracy of the human analyst in determining if an app is malicious.
Accuracy is measured by comparing the result of human-in-the-loop analysis with
the results released by DARPA.
All other factors being equal, 
we found a statistically significant (p $<$ 0.05) decrease in time 
and a statistically significant increase (p $<$ 0.05) in accuracy
with the pushdown-based analyzer.


\paragraph{The challenge suite:}  
Among the apps,  47 are adapted from apps found on the Android market, Contagio~\cite{local:contagio:url}, or the developer's source repository.
A third-party within the APAC project injects malicious behavior into these
apps and uses an anti-diffing tool on apps with larger code bases to make it
difficult to simply diff the application with the original source code.
The remaining five apps are variants of the original 47 apps with different malicious behaviors.
For example, \textit{App1} may leak location information to a malicious website while \textit{App2} may not.
The apps range in size from 18.7 KB to 10 MB, with 11,600 lines of source code in each app on average.

\paragraph{Experiment setup}
The two teams of analysts were given instructions on how to use the tool (both versions of the tool use similar UIs and output forms)
and some warm-up exercises on a couple of examples apps.
Then they were given examples from the DARPA-supplied challenge suite.
The analysts used Anadroid (deployed as a web application on our server) to analyze each app and then 
 record the run time of the analyzer, the total time the human analysis spent investigating the results, and an indication if the analyst felt the app was malicious.

We have made efforts to ensure that other factors remained unchanged so that the only difference was the tool the analyst used to detect malware.
This restriction  can help us gain insight into whether any improvement is made by our new analysis techniques. 

Finally, we compare the analysts results with the DARPA supplied information on whether an app is malicious to check accuracy and run statistical analysis using one-way Analysis of Variance (ANOVA)  to get mean value and p-value of analyzer time,  analyst time, and accuracy.
This allows us to see the statistical results of the experiment on finite-state-based-machine versus pushdown-based control-flow analysis.
This is shown in Table~\ref{tbl:eval-stats2}. 

\begin{table}
\centering
\begin{tabular}{cc|c|c|} 
\cline{3-4} 
& & \textbf{mean} & \textbf{p-value} 
 \\ \cline{1-4}   \cline{1-4}  
 \multicolumn{1}{ |c| }{\multirow{2}{*}{\textbf{Analyzer Time}} } &
 \multicolumn{1}{ |c| }{Finite} &  994 sec &  
 \\ \cline{2-3}
 \multicolumn{1}{ |c|  }{}   &
 \multicolumn{1}{ |c| }{Pushdown} &  560 sec &   0.003 
  \\ \cline{1-4}  
\multicolumn{1}{ |c| }{\multirow{2}{*}{\textbf{Analyst Time}} } &
\multicolumn{1}{ |c| }{Finite} & 1.13 hr &  
\\ \cline{2-3}
\multicolumn{1}{ |c | }{}                        &
\multicolumn{1}{ |c| }{Pushdown} & 0.44 hr &  \textbf{ 0.0} 
\\  \cline{1-4} 
\multicolumn{1}{ |c|  }{\multirow{2}{*}{\textbf{Accuracy}} } &
\multicolumn{1}{ |c| }{Finite} & 71\%&  
\\ \cline{2-3}
\multicolumn{1}{ |c|  }{}                        &
\multicolumn{1}{ |c| }{Pushdown} &  95\%  &  \textbf{0.0005} 
\\  \cline{1-4} 

\end{tabular}
\caption{Comparison of finite-state based  vs. pushdown malware analysis: Pushdown malware analysis yields statistically significant improvements with $p < 0.05$ in both accuracy and analysis time over traditional static analysis method.}
\label{tbl:eval-stats2}
\end{table}

We found that pushdown malware analysis yields statistically significant improvements with $p < 0.05$ in both accuracy and analysis time over traditional static analysis.

\section{Case Studies}\label{sec:cases}

In this section we report case studies of malware detected by Anadroid.
The malicious behavior of the 52 apps provided as part of the APAC challenge
are summarized in Table~\ref{tbl:cases}.
We separate these behaviors into four categories: data leakage, data tampering, denial of service attacks, and other malicious behaviors\footnote{Some apps have more than one malicious behavior.}.

\begin{table*}
\centering
\begin{tabular}{ c c c }

	\textbf{Vulnerabilities} & \textbf{Percentile} & \textbf{Case examples}  
	\\
	Data  leakage   & 57\%  & location, pictures, SMS, ID, etc. ex-filtrated to URL, intents, or predefined local file path.  
	\\
	Data tampering &  10\%&  fill local file system with meaningless data, (recursive) deletion of files
	\\
	DoS attack & 11\% &  inode exhaustion via log, battery drainage (brightness, WiFi, etc.)
	\\
	Other &  28\% & random vibration,  block or intercept  SMS messages

\end{tabular}
\caption{Vulnerabilities summarization}
\label{tbl:cases}
\end{table*}
 
 \textit{Data leakage}  is one of most common, and concerning, malicious behaviors in Android apps~\cite{shuyingliang:Felt:2011:SPSM:MobleMalwareSurvey, shuyingliang:Lu:2012:CCS:CHEX}.
Sensitive data, including location information, an SMS message, or a device ID, is exfiltrated to a third-party host via an HTTP request or Android web component Intent, or to a predefined reachable local file via standard file operations.
This kind of behavior is often embedded in a background Android service component, such as an AsyncTask or a thread, without interfering with the normal functionality of the app.
Anadroid identifies this malicious behavior in 57\% of the 42 malicious Android apps from our test suite that manifest malicious behavior.
We found the taint flow analysis to be more useful than the least permissions analysis in identifying these behaviors, since half of these apps are designed to avoid requesting any permissions.
For instance, instead of requesting the $\syn{ACCESS\_FINE\_LOCATION}$ an app can instead read locations from photos stored on the file system using Exif data and instead of requesting the $\syn{INTERNET}$ the app can use the default Android web view through an $\syn{ACTION\_VIEW}$ Intent, in both cases avoiding the need to explicitly request these permissions. 
 
\textit{Data tampering}, similar to data leakage, can be detected using static taint flow analysis by determining which operations are performed on data.
Malware in this category might corrupt the local file system by overwriting file contents with meaningless data, recursively delete files from the SD card, or delete SMS messages. 
In these real-world apps, exceptions are frequently used, especially around IO operations.
We found that finite-state-based analysis can lead to many spurious execution flows in the control graph when used with EPS.
The pushdown-based model, on the other hand, produces more precise execution flows, which contributes to the sharp decline in analyst time when using Anadroid.

\textit{DoS attack and other malicious behaviors}:  Denial of Service (DoS) on mobile phones exhaust limited resources by intentionally causing the phone to use these resources in an inefficient manner.
For instance, an app might drain the battery by setting brightness to maximum or keeping WiFi on at all times or exhaust file system space by logging every operation to a file.

\textit{Other malicious behaviors} include those that do not leak or tamper with sensitive data but still do not behave the way the app was intended to behave.
For example, a calculator that uses a random number in a calculation rather than the expected number, or blocks SMS messages in the \textit{onReceive} method when a trigger condition is met.
These two categories are more application-dependent and subject to human judgment, by determining if this functionality is too far outside the advertised functionality of the app.
In these scenarios, it is important that the analyzer results not overwhelm analysts. 

Anadroid cannot determine precisely whether the application is malicious or not by itself.
Instead, it identifies suspicious application behavior and uses analyst-supplied predicates to help search analysis results for locations of interest to the analyst.

\section{Related work} \label{sec:related}
 \paragraph{Static analysis for Java programs}
 
 Precise and scalable context-sensitive pointer analysis for Java (object-oriented) programs
 has been an open problem for decades. Remarkably, 
 a large bulk of the previous literature focused on finite-state abstractions for Java programs,
     \ie~$k$-CFA, limited object sensitivity, and their variants.
 In work that addresses exception flows~\cite{Robillard:2003:evolution-exception,local:Jo:2002:UncaughtException,Ryu:2001:MultiThreadExceptions,Leroy:2004:TypeBasedUncaughtExceptions}, the analysis is often based on
  context-insensitivity or limited context-sensitivity, which means they cannot differentiate the contexts 
  where an exception is thrown or precisely determine which handlers can handle an exception.
   
Spark~\cite{shuyingliang:Lhotak:2003:CC:SPARK} and Paddle~\cite{shuyingliang:Lhotak:2008:BDDOO} both use imprecise exception analysis. 
 Soot~\cite{local:soot:url} also uses a separate exception analysis implemented by  Fu~\etal{}~\cite{shuyingliang:Fu:2007:Exception-Chain-Analysis} which is not based on pointer analysis and not integrated into the tool.
 Bravenboer and Smaragdakis~\cite{Bravenboer:2009:Exceptions}
   propose joining points-to analysis and exception flow analysis to improve 
   precision and analysis run time in their Doop framework~\cite{Bravebboer:2009:declare-pointsto}.
   They have conducted extensive comparison of different options for polyvariance.
   It provides a more precise and efficient exception-flow analysis than Spark, Paddle, and Soot, with respect of points-to 
  and exception-catch links with respect to the metric used in~\cite{Fu:2005:rubust-java-server-apps, Bravebboer:2009:declare-pointsto}. 
  IBM Research's WALA is a static analysis library designed to support different pointer analysis configurations.
 The points-to analyses of WALA can compute which exceptions a method can throw, 
 but does not guarantee precise matches between exceptions and their corresponding handlers.

 \paragraph{ CFL- and pushdown-reachability techniques}
 Earl~\etal~\cite{Earl:2012:IPDCFA}
 develop a pushdown reachability algorithm
 suitable for pushdown systems, which essentially draws on CFL- and pushdown-reachability
  analysis~\cite{mattmight:Bouajjani:1997:PDA-Reachability,mattmight:Kodumal:2004:CFL,mattmight:Reps:1998:CFL,mattmight:Reps:2005:Weighted-PDA}.
 We modify their traditional CESK machine to handle object-oriented programs and extend it to analyze exceptions.
This allows us to apply their algorithm directly to our analysis.
 CFL-reachability techniques have also been used to compute classical
 finite-state abstraction CFAs~\cite{mattmight:Melski:2000:CFL} and
 type-based polymorphic control-flow
 analysis~\cite{mattmight:Rehof:2001:TypeBased}.
 These analyses should not be confused with pushdown control-flow
 analysis, which is a fundamentally different kind of CFA.
 
 \paragraph{Malware detection for Android applications}
 Several analyses have been proposed for Android malware detection.
 
 Dynamic taint analysis has been applied to identify security vulnerabilities at run time in Android applications.
 TaintDroid~\cite{shuyingliang:Enck:2010:TaintDroid} dynamically tracks the flow of sensitive information and looks for confidentiality violations.
 IPCInspection~\cite{shuyingliang:Felt:2011:IPCInspection}, QUIRE~\cite{shuying:Dietz:2011:Quire}, and XManDroid~\cite{shuyingliang:Bugiel:2012:XManDroid}  are designed to prevent privilege-escalation, 
 where an application is compromised  to provide sensitive capabilities to other applications. 
 The vulnerabilities introduced by interapp communication is considered future work.
 However, these approaches typically ignore implicit flows raised by control structures in order to reduce run-time overhead. 
 Moreover, dynamically executing all execution paths of these applications to detect potential information leaks is impractical. 
 The limitations make these approaches inappropriate for computing information flows for all submitted applications. 
 
  Woodpecker~\cite{shuyingliang:Grace:2012:Woodpecker} uses traditional data-flow analysis to find possible capability leaks. Comdroid~\cite{shuyingliang:Chin:2011:Comdroid} targets vulnerabilities related to interapp communications. However, it does not perform deep program analysis as Anadroid does, and this results in high false positive rates.
  SmartDroid~\cite{shuyingliang:Zheng:2012:SPSM:SmartDroid} targets finding complex UI triggers and paths that lead to sensitive sinks. It addresses imprecision of static analysis by combining dynamic executions to filter out infeasible paths at run time.
 CHEX~\cite{shuyingliang:Lu:2012:CCS:CHEX} focuses on detecting \textit{component hi-jacking}  by augmenting existing analysis framework using app-splitting to handle Android's multiple entry points.  Our tool takes a significantly different approach from it (and other finite-state-based static analysis tools) in three aspects. (1) We use pushdown flow analysis that handles traditional control-flow and exception flows precisely and efficiently.  (2) Our Entry Point Saturation technique is sound, and we are able to detect interleaving execution of multiple entry points while CHEX handles only permutations of multi-entry points. (3) Our tool enables human-in-the-loop analysis by allowing the analyst to supply predicates for the analyzer allowing it to highlight inspection of deeply disguised malware.

 
Jeon~\etal{}~\cite{shuyingliang:Jeon:2012:DAM} proposes enforcing a fine-grained permission system. It limits access to resources 
 that could normally be accessed by one of Android's default permissions.
 Specifically, the security policy uses a white list to determine which resources an app can use and a black list to deny access to resources.
 In addition,  strings potentially containing URLs are  identified by pattern matching and constant propagation is used to infer more specific Internet permissions. 
Grace~\etal{}~\cite{shuyingliang:grace:2012:NDSS:capability} have also identified unprivileged malicious apps that can exploit permissions on protected resources through a privileged agent (or app component in our test suite) that does not enforce permission checks.  Anadroid can also identify this malicious behavior.
 
Stowaway~\cite{shuyingliang:Felt:2011:Stowaway} is a static analysis tool 
identifying whether an application requests more permissions than it actually uses. 
PScout~\cite{shuyingliang:Au:2012:CCS:PScout} aims for a similar goal, but produces more precise and  fine-grained mapping from APIs to permissions.
Our least permission report uses the PScout permission map as Anadroid's database.
However, they use a different approach, adapting testing methodology to test applications and identify APIs that require permissions,
while our approach annotates APIs with permissions and statically analyzes all executable paths. 

%
\section{Conclusion}\label{sec:concl}

In this paper, we address two challenges in static malware detection for Android apps: the fundamental challenge of analyzing object-oriented programs and the Android domain specific challenge of asynchronous multi-entry points. We address the first challenge using pushdown control flow analysis (PDCFA) to precisely analyze both traditional control flows and exception flows. The second challenge is addressed  via entry-point saturation (EPS) that when integrated with PDCFA serves as the basis for our analysis engine.  We demonstrate a malware analyzer built on this engine, adding pushdown taint flow and least permissions analysis. We describe Anadroid, a generic analysis framework for Dalvik-bytecode that enables human-in-the-loop analysis by accepting user-supplied predicates  to search analysis results for detailed inspection. We compare the new analyzer with a traditional finite-state analyzer using a test suite released by DARPA APAC project.  We find that PDCFA together with EPS yields statistically significant improvements in both accuracy and analysis time over traditional static analysis methods. Our implementation  is publicly available: \url{github.com/shuyingliang/pushdownoo}.

\vfill\eject
\section{Acknowledgments}

This material is based on research sponsored by DARPA under agreement
number FA8750-12- 2-0106. The U.S. Government is authorized to
reproduce and distribute reprints for Governmental purposes
notwithstanding any copyright notation thereon.



\end{document}